\documentclass[nato,noid,numreferences]{crckbked}

\usepackage{graphicx}
\usepackage{epsf}

\begin{document}

\begin{opening}
\title{ MATRIX MODELS ON THE FUZZY SPHERE}

\author{BRIAN P. DOLAN}
\institute{Department of Mathematical Physics,\\ National
University of Ireland, Maynooth, Ireland}

\author{DENJOE O'CONNOR}
\institute{School of Theoretical Physics,\\ Dublin Intistitute of
Advanced Study, Dublin, Ireland}
\author{and PETER PRE\v{S}NAJDER}
\institute{Department of Theoretical Physics,\\ Comenius
University, Bratislava, Slovakia}

\begin{abstract}

Field theory on a fuzzy noncommutative sphere can be considered as
a particular matrix approximation of field theory on the standard
commutative sphere. We investigate from this point of view the
scalar $\phi^4$ theory. We demonstrate that the UV/IR mixing
problems of this theory are localized to the tadpole diagrams and
can be removed by an appropiate (fuzzy) normal ordering of the
$\phi^4$ vertex. The perturbative expansion of this theory reduces
in the commutative limit to that on the commutative sphere.
\end{abstract}

\end{opening}

\section{Introduction}

The field-theoretical models possessing a finite number of field
modes can be presented as matrix models. To this class of models
belong, e.g., the models on a finite cubic lattice subjected to
some periodic boundary conditions. Another class of such models
represent models on a fuzzy sphere, \cite{GKP1}-\cite{bal1}. The
basic idea behind fuzzy sphere is quite simple:

(i) The standard sphere $S^2$ with radius $R$ is a co-adjoint
orbit possessing a Poisson bracket, and it can be considered as a
particular "phase-space" with finite volume $4\pi R^2$;

(ii) After "quantization" (deformation), this "phase-space" is
effectively divided to cells of the size $\sim\rho^2$ ($\rho$ is a
deformation parameter), each representing one field mode. Thus,
the total number of field modes is $N\sim R/\rho$. The resulting
"quantum" space is known as the fuzzy sphere $S^2_F$, for details
see \cite{mad}.

In the framework of lattice models it is of great interest to
investigate their properties in the continuum limit when, in a
proper sense, the lattice spacing $a$ approaches zero: $a\to 0$.
For models on a fuzzy sphere the commutative limit $\rho\to 0$ (or
equivalently, $N\to\infty$) plays a similar role. A simple scalar
field model on a fuzzy sphere with $\phi^4$ interaction was
proposed in \cite{GKP1}.

A potential problem for this program has emerged due to the
phenomenon of UV/IR mixing, \cite{MvRS}. The problem was discussed
on a noncommutative Moyal space, where fields possess an infinite
number of modes, and consequently a regularization procedure is
needed. The phenomenon appears to be generic for noncommutative
spaces.

It was pointed out in \cite{vai} that the UV/IR mixing is present
in the $\phi^4$ model with naive action introduced in \cite{GKP1}.
This work was followed by \cite{CMS}, where the one-loop
contribution to the two-point vertex function was calculated
explicitely, and it was shown that in the commutative limit there
is a finite correction to the expected commutative contribution.
Moreover, in the planar limit it incorporates the UV/IR mixing
singularity of the Moyal space.

The implications of this result are very serious for the program
of using matrix model approximation to continuum field theories.
Recently, in \cite{BDP} we found that there is a quite natural
solution and that the problem disappears when the interaction term
is properly (fuzzy) normal ordered. The resulting action is
therefore the correct starting point for fuzzy lattice
approximations.

In the next section we briefly describe the Euclidean scalar field
with $\phi^4$ interaction on a commutative sphere. Then we present
its fuzzy version and formulate all essential results found in
\cite{BDP}. The last section contains a brief summary.\\

\section{Model on a standard commutative sphere.}

We consider a real $\Phi(\vec{n})$ defined on a usual sphere $S^2$
with radius $R$: $\vec{n}=(n_1, n_2,n_3)$, ${\vec{n}}^2=R^2$. The
Euclidean field action for $\Phi(\vec{n})$ with $\Phi^4(\vec{n})$
interaction is given as \begin{equation} S[\Phi]\ =\ \int dn\,
[{1\over 2}({\cal L}_i\Phi)^2 + {r\over 2}\Phi^2 + {\lambda\over
4!}\Phi^4]\ ,\end{equation} where $dn=\sin\theta d\theta d\varphi$
is the standard measure on $S^2$ normalized to $4\pi$, and ${\cal
L}_i=i\varepsilon_{ijk}n_j\partial_{n_k}$, $i,j,k=1,2,3$, are the
usual generators of rotations.

The Euclidean QFT can be formulated, e.g., in terms of functional
integrals. If $F[\Phi]$ is some functional depending on field
$\Phi=\Phi(\vec{n})$, then its quantum average is defined by
\begin{equation} \langle F[\Phi]\rangle\ =\ \int D\Phi\,
e^{-S[\Phi]}F[\Phi]\ .\end{equation} Here the functional measure
$D\Phi$ is normalized by $\langle 1\rangle =1$. It can be shown
that the model specified by (1) and (2) can be formulated
rigorously, and moreover, that it possesses the perturbative
expansion in $\lambda$ which is Borel summable.

Therefore, we can restrict ourselves to the standard perturbation
expansion of the model based on on (1). It is well known that it
contains only finite Feynman diagrams except the tadpole diagram
which contribution \begin{equation} g_L(r)\ =\ {1\over
4\pi}\sum_{l=0}^L {2l+1\over l(l+1)+r}\end{equation} is
logharithmically divergent in the cut-off parameter $L$. In
interaction picture the interaction term in the action is normal
ordered \begin{equation} S_{int}[\Phi]\ =\ \int dn\, {\lambda\over
4!} :\Phi^4:\ ,\end{equation} where the normal ordering is
specified by \begin{equation} :\Phi^4:\ =\ \Phi^4\, -\,
12g(L,t)\Phi^2\ .
\end{equation} The parameters $r$ and $t$ are related by $r=t-
(\lambda /2)g_L(t)$. This procedure eliminates all tadpole
diagrams, and the resulting perturbative series contains only
finite Feynman diagrams.

\section{Model on the fuzzy sphere}

The standard sphere $S^2$ with radius $R$ is fully characterized
by its Cartesian coordinates $n_i$, $i=1,2,3$, satisfying
relations \begin{equation} [n_i,n_j]=n_in_j-n_jn_i=0\ ,\ \sum_i
n^2_i=R^2\ .\end{equation} In the noncommutative case one replaces
these commuting coordinates by operators ${\hat n}_i$, $i=1,2,3$,
satisfying relations \begin{equation} [{\hat n}_i,{\hat n}_j]=
{\hat n}_i{\hat n}_j-{\hat n}_j{\hat n}_i=i\rho\varepsilon_{ijk}
{\hat n}_k\ ,\ \sum_i{\hat n}^2_i=R^2\ .\end{equation} If
$R^2/\rho^2=L(L+1)$ with $L$ positive integer, then ${\hat n}_i$,
$i=1,2,3$, can be realized as $(L+1)\times (L+1)$ matrices (in
fact, conditions (7) determine the spin $s=L/2$ representation of
$SU(2)$ group). The scalar field is now a polynomial in ${\hat
n}_i$, i.e. a general $(L+1)\times (L+1)$ matrix ${\hat\Phi}= \Phi
({\hat n})$.

The fuzzy analogs ${\hat{\cal L}}_i$, $i=1,2,3$, of the
differential operators ${\cal L}_i$, act on an arbitrary matrix
${\hat f}=f({\hat n})$ as follows: \begin{equation} {\hat{\cal
L}}_i{\hat f}\ =\ [{\hat L}_i,{\hat f}] \ ,\  {\hat L}_i=
\rho^{-1}{\hat n}_i\ ,\ i=1,2,3\ .\end{equation} In the fuzzy
case, the integral over $S^2$ is replaced by the normalized trace
\begin{equation} \int dn\, f(\vec{n})\ \to\ {4\pi\over L+1}\,
{\rm Tr} f({\hat n})\ .\end{equation} In the commutative limit
$\rho\to 0$, i.e. $L\to\infty$, eqs. (8) and (9) reduce,
respectively, to the standard definitions of generators of
rotations and the usual integral over $S^2$.

The modifications indicated in (8) and (9) lead to the na\"{\i}ve
fuzzy field action proposed in \cite{GKP1}: \begin{equation}
S_L[{\hat\Phi}]\ =\ {4\pi\over L+1}\, {\rm Tr} [{1\over
2}({\hat{\cal L}}_i{\hat\Phi})^2 + {r\over 2}{\hat\Phi}^2 +
{\lambda\over 4!}{\hat\Phi}^4]\ .\end{equation} Of course, this
gives a tadpole contributions to the pertubation series, which
indeed, are finite since the number of field modes is finite.

However, in the noncommutative case the permutation symmetry of
the vertex contributions, originated from $\int dn\Phi^4$, is
reduced the cyclic symmetry of the fuzzy vertex $4\pi(L+1)^{-1}\,
{\rm Tr}{\hat\Phi}^4$. This yealds two independent tadpole
contributions: instead of 12 tadpole subtractions in (5), we have
now 8 subtractions of the planar tadpole diagram and 4
subtractions of the non-planar one. The planar diagram gives the
same contribution as the tadpole in the commutative case, whereas
the nonplanar possesses a correction. Therefore, the correct fuzzy
interaction is, \cite{BDP}: \begin{equation} S_{int}[{\hat\Phi}]\
=\ {4\pi\over L+1}\, {\rm Tr} :{\hat\Phi}^4:]\ ,\end{equation}
with the normal ordering defined as follows: \begin{equation} {\rm
Tr}:{\hat\Phi}^4:\ =\ {\rm Tr}\left[ {\hat\Phi}^4\, -\,
12\sum_{lm}{{\hat\Phi}{\hat Y}_{lm}{\hat Y}_{lm}{\hat\Phi}\over
l(l+1)+t} \, +\, 2\sum_{lm}{[{\hat\Phi},{\hat Y}_{lm}]^\dagger
[{\hat\Phi},{\hat Y}_{lm}]\over l(l+1)+t} \right]\ .\end{equation}
Here, ${\hat Y}_{lm}$ are fuzzy analogs of spherical functions
determined by the conditions: \begin{equation} {\hat{\cal
L}}^2_i{\hat Y}_{lm}= l(l+1){\hat Y}_{lm}\ ,\ {\hat{\cal
L}}_3{\hat Y}_{lm}=m{\hat Y}_{lm}\ ,\ {4\pi\over L+1}{\rm Tr}{\hat
Y}^\dagger_{lm}{\hat Y}_{l'm'}=\delta_{ll'}\delta_{mm'}\
.\end{equation} The middle term in (12) is the usual tadpole
subtraction and corresponds to a normal ordering in the
commutative vertex. The last term is the additional noncommutative
subtraction that is necessary to obtain the correct commutative
limit.

\section{Conclusions}

The correct matrix model that represents a lattice regularization
of the commutative theory is given by the field action with the
interaction term given by (11) and (12). It can be put into the
form \begin{equation} {\tilde S}_L[{\hat\Phi}]\ =\ {4\pi\over
L+1}\, {\rm Tr} [{1\over 2}{\hat\Phi}({\hat{\cal L}}^2 +
{\lambda\over 2}Q_L({\hat{\cal L}}^2)+t-g_L(t)){\hat\Phi} +
{\lambda\over 4!}{\hat\Phi}^4]\ .\end{equation} The term $(\lambda
/2)g_L(t)$, given by the middle term in (12), corresponds to the
usual tadpole subtraction (the parameter r is replaced by
$t-(\lambda /2)g_L(t)$, exactly, like in the commutative theory).
The term $(\lambda /2)Q_L({\hat{\cal L}}^2)$ is given by the
manifestly positive last term in (12). It interpreted as the
momentum dependent wave-function renormalization since
$Q_L({\hat{\cal L}}^2)$ is a power series in ${\hat{\cal L}}^2$
which starts at ${\hat{\cal L}}^2$, and therefore we could have
written (see, \cite{CMS}, \cite{BDP}): \[ {\hat{\cal
L}}^2+{\lambda\over 2}Q_L({\hat{\cal L}}^2)\ =\ {\hat{\cal
L}}^2Z_L({\hat{\cal L}}^2)\ .\] This term exactly cancels the
unwanted momentum dependent quadratic terms in the effective
action arising from non-planar diagrams in the fuzzy theory. This
cancellation then guarantees that the continuum limit of this
theory is the standard scalar $\phi^4$ on the sphere.

The modified action (14) can serve for the non-perturbative
definition of the fuzzy quantum averages \begin{equation} \langle
F[{\hat\Phi}]\rangle_L\ =\ \int D_L{\hat\Phi}\, e^{-{\tilde
S}_L[{\hat\Phi}]}F[{\hat\Phi}]\ .\end{equation} Here,
$D_L{\hat\Phi}=N_L\Pi_{lm}dc_{lm}$ is a finite-dimensional measure
associated to arbitrary field configurations \begin{equation}
{\hat\Phi}\ =\ \sum_{l=0}^L\sum_{m=-L}^{+L}\, c_{lm}{\hat Y}_{lm}\
,\ c_{l,-m}=c^*_{l,m}\ -\ {\rm complex}\ ,\end{equation} with the
normalization constant $N_L$ fixed by $\langle 1\rangle_L =1$. The
action ${\tilde S}_L[{\hat\Phi}]$ in (15) guarantees the correct
commutative limit $L\to\infty$ of quantum averages  $\langle
F[{\hat\Phi}]\rangle_L$ for any fixed field functional
$F[{\hat\Phi}]$ not depending explicitely on L.

It would be desirable to investigate in the same spirit the
$\phi^4$ theory in four-dimensional fuzzy spaces, e.g. as a
suitable canditate could serve the fuzzy space $S^2_F\times
S^2_F$. However, in this case, the problem will be more severe
since there will be additional residual non-local differences for
two- and four-point functions. It will be therefore more difficult
to establish the model which will reproduce the commutative limit.

\end{document}